\documentclass[10pt]{article}
\usepackage{amsmath}
\usepackage{amssymb}
\usepackage{graphicx}
\usepackage{longtable}
\usepackage{multirow}
\usepackage{multicol}
\usepackage{textcomp}
\usepackage[utf8]{inputenc}
\usepackage[british]{babel}
\usepackage{float}
\usepackage{gensymb}
\usepackage{amsfonts}
\usepackage{cite}
\usepackage{makecell}
\usepackage{caption}
\usepackage{epstopdf}
\usepackage[export]{adjustbox}
\usepackage[paperwidth=595.44pt,
paperheight=841.68pt,top=72pt,right=72pt,bottom=72pt,
left=72pt]{geometry
}
\begin{document}
\title{Controlled Quantum Teleportation of Superposed Coherent State using GHZ Entangled Coherent State
}
\date{}
\maketitle
\begin{center}
Ravi Kamal Pandey\textsuperscript{*},
Ranjana Prakash\textsuperscript{**},
Hari Prakash\textsuperscript{***},
\\
\bigskip
\ Physics Department, University of Allahabad, Allahabad- 211002, India.
\\
E-mail:* ravikamalpandey@gmail.com\\
 ** prakash\_ranjana1974@rediffmail.com\\
*** prakash\_hari123@rediffmail.com
\end{center}

\begin{abstract}
Controlled quantum teleportation of superposed coherent states using GHZ entangled 3-mode coherent states is studied. Proposed scheme can be implemented experimentally using linear optical components such as a symmetric lossless beam splitter, two phase-shifters and two photon counters. Fidelity is found close to unity for appreciable mean number of photons in coherent states and is 0.99 for mean photon number equal to two.
\end{abstract}
\section{Introduction}
Entangled systems shows some bizzare correlations between its sub-systems, that has no classical counterparts. These have proven to be a useful physical resource for performing many useful tasks such as quantum computing \cite{deutsch1985quantum}, quantum teleportation \cite{bennett1993teleporting}, super dense coding \cite{bennett1992communication}, remote state preparation \cite{pati2000minimum,lo2000classical}, quantum cryptography \cite{ekert1991quantum} etc. Bennet et al \cite{bennett1993teleporting} gave protocol for teleporting from one party, say, Alice to another distant party, Bob, the quantum state of a two-level system spanning a two-dimensional Hilbert space, using a two-partite entangled resource, the particles being shared between Alice and Bob. This idea called standard quantum teleportation (SQT) was extended to controlled quantum teleportation (CQT) involving three-partite entangled resource, the addition particle being sent to a third party, say, Charlie, who acts as a supervisor and controls the process \cite{karlsson1998quantum}. The idea was further extended in teleportation of continuous variables i.e. sates living in infinite dimensional Hilbert space\cite{vaidman1994teleportation}. The theoretical schemes were further realized in many experiments for teleportation of discrete as well as continuous variables \cite{polkinghorne1999continuous,bouwmeester1997experimental,kim2001quantum}.\\ 
Coherent state of radiation field, $|\alpha\rangle=\sum_{n=0}^{\infty} e^{-1/2|\alpha|^{2}}\frac{(\alpha)^{n}}{n!}|n\rangle$  \cite{glauber1963coherent}, with $\alpha$ any complex number, are eigenstates of photon annihilation operator and are closest to a classical noiseless field. The coherent states are in general non-orthogonal, in fact for two coherent state $|\alpha\rangle$ and $|\beta\rangle$ their overlap $|\langle\alpha|\beta\rangle|^{2}=$ exp$(-|\alpha-\beta|^{2})$. Orthogonal states can be obtained by considering even and odd coherent states, where
\begin{equation}\label{eq-1}
|EVEN,\alpha\rangle=[\sqrt{2(1+x^{2})}]^{-1}(|{\alpha}\rangle+|-{\alpha}\rangle) \mbox{ and } |ODD,\alpha\rangle=[\sqrt{2(1-x^{2})}]^{-1}(|{\alpha}\rangle-|-{\alpha}\rangle),
\end{equation}
are well known schr\"{o}dinger cat states \cite{dodonov1974even}. It is customary to write information as, $a|\alpha\rangle+b|-\alpha\rangle$ or as, $A_{+}|EVEN,\alpha\rangle+A_{-}|ODD,\alpha\rangle$. This allows for the possibility of encoding and manipulating information in superposed coherent state\cite{ralph2003quantum,sanders2012review}. The quantum resource required for the teleportation of information encoded in the superposition of coherent state can be generated by using a nonlinear Mach Zhender interferometer, in which one of the arms of interferometer has a non-linear medium of second order non-linearity \cite{sanders1992entangled}. Entangled coherent states (ECS) has been shown to be more robust against de-coherence due to photon absorption than standard bi-photonic entangled states which makes it an important candidate for using it as quantum channel \cite{park2010entangled,hirota2001entangled}.\\ Van enk et al \cite{van2001entangled} proposed a scheme of standard quantum teleportation of one qubit of information encoded in superposition of coherent states (SCS) $|\alpha\rangle$ and $|-\alpha\rangle$ 
  \begin{equation} \label{eq-2}
  	|I\rangle=\epsilon_{+}|\alpha\rangle+\epsilon_{-}|-\alpha\rangle,
  \end{equation}
 whose normalization demands $|\epsilon|_{+}^{2}+|\epsilon|_{-}^{2}+2x^{2}Re(\epsilon_{+}^{\star}\epsilon_{-})=1$ 
using maximally entangled bi-partite ECS,
\begin{equation} \label{eq-3}
|\psi\rangle_{1,2}=[2(1-x^{4})]^{-1/2}(|\alpha,\alpha\rangle-|-\alpha,-\alpha\rangle)_{1,2}
\end{equation}
with success probability $1/2$. Wang et al. \cite{wang2001quantum} extended this idea in teleporting ECS also with same success probability. N Ba An \cite{an2003teleportation} discussed a scheme of quantum teleportation of superposed coherent state (SCS) within a network, also with success probability $1/2$. Prakash et al. \cite{prakash2007improving} obtained almost perfect success in teleportation for appreciable mean number of photons in coherent states, by altering the photon counting scheme. The scheme was further used extensively for improving various quantum information processing task involving ECS \cite{mishra2010teleportation,mishra2015long,prakash2009entanglement,
prakash2009quantum,prakash2009swapping,prakash2010almost,prakash2010improving,
prakash2011quantum,prakash2013quantum,Prakash2019}.
CQT of the state of eq. \eqref{eq-2} using four mode cluster type ECS \cite{an2009cluster} was proposed by liu et al \cite{liu2011controlled} and reported success probability 1/2. To best of our knowledge nobody has studied the simpler CQT using only three mode ECS, which is surprising. In this paper, we present a simpler scheme of CQT of state eq. \eqref{eq-2}, using GHZ type ECS. It is shown that CQT with success probability one and fidelity of teleportation, defined as the overlap between teleported and information state, approaching unity for significant mean photons in the concerned coherent state and hence an almost perfect CQT is possible.
\section{CQT of SCS}
Let us consider two remote partners Alice and Bob. Alice has an information encoded in SCS as given by eq. \eqref{eq-1} (in mode, say, 0), which she wanted to teleport to Bob. Using even and odd coherent states, one can have Bloch space representation of the information state,\begin{equation} \label{eq:4}
|I\rangle=A_{+}|EVEN,\alpha\rangle+A_{-}|ODD,\alpha\rangle
\end{equation}
where one may define angles $\theta$ and $\phi$ by $A_{+}=cos(\theta/2)$ and $A_{-}=e^{i\phi}sin(\theta/2)$, satisfying the normalization condition, $|A_{+}|^{2}+|A_{-}|^{2}=1$. One can readily go from the non-orthogonal eq. \eqref{eq-1} to orthogonal  representation $A_{+}|EVEN,\alpha\rangle+A_{-}|ODD,\alpha\rangle$ or vive-versa, of the information using the interrelationship $A_{\pm}=\sqrt{(1\pm x^{2})/2}(\epsilon_{+}+\epsilon_{-})$ and $\epsilon_{\pm}=(A_{+}/\sqrt{(1+ x^{2})/2})\pm (A_{-}/\sqrt{(1- x^{2})/2})$ . To enhance the security of the teleportation protocol, we introduce one more partner, Charlie, playing the role of a controller. We consider a 3-mode maximally entangled  GHZ type ECS
\begin{equation} \label{eq-5}
|G\rangle_{2,3,4}=N_{G}(|\alpha,\alpha,\alpha\rangle_{2,3,4}+|\alpha,\alpha,\alpha\rangle_{2,3,4}),
\end{equation}
as the quantum channel, where $N_{G}=[2(1-x^{6})]^{-1/2}$ is the normalization parameter. Out of modes 2, 3 and 4, 2 goes to Alice, 3 to Bob and 4 to Charlie. The state of the system consisting of modes 1, 2, 3 and 4 is given by,
\begin{equation} \label{eq-6}
|\psi\rangle_{1,2,3,4}=N_{G}[\epsilon_{+}(|\alpha,\alpha,\alpha,\alpha\rangle_{1,2,3,4}-|\alpha,-\alpha,-\alpha,-\alpha\rangle_{1,2,3,4})$$$$
-\epsilon_{-}(|-\alpha,\alpha,\alpha,\alpha\rangle_{1,2,3,4}-|-\alpha,-\alpha,-\alpha,-\alpha\rangle_{1,2,3,4})].
\end{equation}
Modes 1 and 2 are with Alice. She passes mode 2 through a phase-shifter which converts states $|\beta\rangle_{2}$ to $|-i\beta\rangle_{5}$. Alice mixes using a symmetric lossless beam-splitter the states $|\beta,\gamma\rangle_{1,3}$ in modes 1 and 3 to states $|\dfrac{\beta+i\gamma}{\sqrt{2}},\dfrac{\gamma+i\beta}{\sqrt{2}}\rangle_{6,7}$. The mode 6 is then passed through a phase-shifter which changes state $|\delta\rangle_{6}$ to $|-i\delta\rangle_{7}$. The combination of two phase shifters and beam-splitter thus changes state $|\alpha,\beta\rangle_{1,2}$ to the state $|\dfrac{\alpha+\beta}{\sqrt{2}},\dfrac{\alpha-\beta}{\sqrt{2}}\rangle_{6,8}$. Using this prescription, we can write $|\psi\rangle_{1,2,3,4}$ in terms of states $\pm |\sqrt{2}\alpha\rangle_{6}$, $|0\rangle_{6}$, $\pm |\sqrt{2}\alpha\rangle_{8}$, $|0\rangle_{8}$, $\pm |\alpha\rangle_{3}$ and $\pm |\alpha\rangle_{4}$ and get
\begin{equation}\label{eq-7}
|\psi\rangle_{6,8,3,4}=N_{G}[\epsilon_{+}(|\sqrt{2}\alpha,0,\alpha,\alpha\rangle_{6,8,3,4}-|0,\sqrt{2}\alpha,-\alpha,-\alpha\rangle_{6,8,3,4})$$$$-\epsilon_{-}(|0,-\sqrt{2}\alpha,\alpha,\alpha\rangle_{6,8,3,4}-|-\sqrt{2}\alpha,0,-\alpha,-\alpha\rangle_{6,8,3,4})].
\end{equation}
Prakash et al. \cite{prakash2007improving} showed that that coherent state $|\alpha\rangle$  can be written as superposition of states having (i) no photons, (ii) non-zero even number of photons and (iii) odd number of photons and wrote
\begin{equation} \label{eq-8}
|\pm\sqrt{2}{\alpha}\rangle=x|0\rangle+\frac{1-x^2}{\sqrt{2}}|{NZE,\sqrt{2}\alpha}\rangle\pm\sqrt{\frac{1-x^4}{2}}|{ODD,\sqrt{2}\alpha}\rangle
 \end{equation}
 where, $|NZE,\sqrt{2}\alpha\rangle=[\sqrt{2}(1-x^{2})]^{-1}(|\sqrt{2}\alpha\rangle+|-\sqrt{2}\alpha\rangle-2x|0\rangle)$ is the state containing non-zero even number of photons. Using this we can write eq. \eqref{eq-7} in orthonormal basis $|0\rangle$, $|NZE,\alpha\rangle$ and $|ODD,\alpha\rangle$ as, 
\begin{equation}\label{eq-9}
|\psi\rangle_{6,8,3,4}=N_{G} \{ x\sqrt{2(1-x^{2}})[|0,0\rangle_{6,8}|EVEN,\alpha\rangle_{3}(A_{+}|ODD,\alpha\rangle_{4})+
|0,0\rangle_{6,8}|ODD,\alpha\rangle_{3}(A_{+}|EVEN,\alpha\rangle_{4})]
$$$$
+\dfrac{1}{2}(1-x^{2})\sqrt{1+x^{2}}[|NZE,\sqrt{2}\alpha\rangle_{6}|0\rangle_{8}|EVEN,\alpha\rangle_{3}(\sqrt{\dfrac{1+x^{2}}{1-x^{2}}}A_{-}|EVEN,\alpha\rangle_4+\sqrt{\dfrac{1-x^{2}}{1+x^{2}}}A_{+}|ODD,\alpha\rangle_4)
$$$$
+|0\rangle_{6}|NZE,\sqrt{2}\alpha\rangle_{8}|EVEN,\alpha\rangle_{3}(-\sqrt{\dfrac{1+x^{2}}{1-x^{2}}}A_{-}|EVEN,\alpha\rangle_4+\sqrt{\dfrac{1-x^{2}}{1+x^{2}}}A_{+}|ODD,\alpha\rangle_4)]
$$$$
+\dfrac{(1-x^{2})^{3/2}}{2}[|NZE,\sqrt{2}\alpha\rangle_{6}|0\rangle_{8}|ODD,\alpha\rangle_{3}(A_{+}|EVEN,\alpha\rangle_4+A_{-}|ODD,\alpha\rangle_4)+|0\rangle_{6}|NZE,\sqrt{2}\alpha\rangle_{8}|ODD,\alpha\rangle_{3} 
$$$$
(A_{+}|EVEN,\alpha\rangle_4-A_{-}|ODD,\alpha\rangle_4)]+\dfrac{1}{2}(1+x^{2})\sqrt{1-x^{2}}[|ODD,\sqrt{2}\alpha\rangle_{6}|0\rangle_{8}|EVEN,\alpha\rangle_{3}
$$$$
(A_{+}|EVEN,\alpha\rangle_4+A_{-}|ODD,\alpha\rangle_4)+|0\rangle_{6}|ODD,\sqrt{2}\alpha\rangle_{8}|EVEN,\alpha\rangle_{3}(-A_{+}|EVEN,\alpha\rangle_4+A_{-}|ODD,\alpha\rangle_4)]+
$$$$
\dfrac{1}{2}(1-x^{2})\sqrt{1+x^{2}}[|ODD,\sqrt{2}\alpha\rangle_{6}|0\rangle_{8}|ODD,\alpha\rangle_{3}
(\sqrt{\dfrac{1+x^{2}}{1-x^{2}}}A_{-}|EVEN,\alpha\rangle_4
+\sqrt{\dfrac{1-x^{2}}{1+x^{2}}}A_{+}|ODD,\alpha\rangle_4)
$$$$
+|0\rangle_{6}|ODD,\sqrt{2}\alpha\rangle_{8}|ODD,\alpha\rangle_{3}
(\sqrt{\dfrac{1+x^{2}}{1-x^{2}}}A_{-}|EVEN,\alpha\rangle_4-\sqrt{\dfrac{1-x^{2}}{1+x^{2}}}A_{+}|ODD,\alpha\rangle_4)]\}
\end{equation}

\section{Analysis of Photon Count Results}
\begin{figure}[!htb]
   \begin{minipage}{0.48\textwidth}
     \centering
     \includegraphics[width=.9\linewidth, trim=0 .2cm .65cm 0, clip]{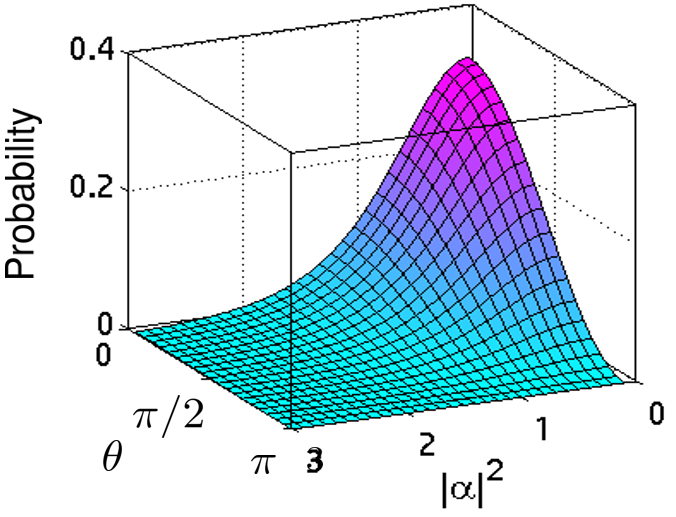}
     \caption{Variation of probability of occurrance with respect to information parameter $\theta$ and average photon number in the coherent state $|\alpha|^{2}$ for cases 1 and 2.}\label{Fig:fig-1}
   \end{minipage}\hfill
   \begin{minipage}{0.48\textwidth}
     \centering
     \includegraphics[width=.9\linewidth, trim=0 1.4cm .4cm 0, clip]{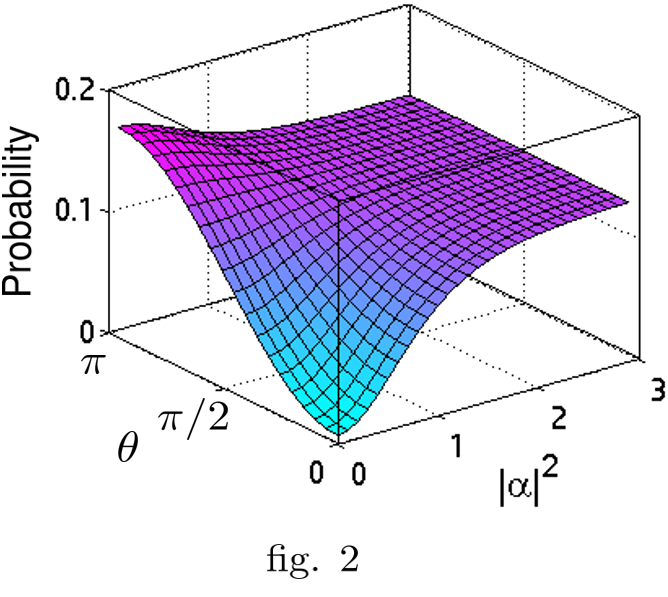}
     \caption{Variation of probability of occurrance with respect to information parameter $\theta$ and average photon number in the coherent state $|\alpha|^{2}$ for cases 3, 5, 8 and 10.}\label{Fig:fig-2}
   \end{minipage}
\end{figure}

Alice performs photon count (PC) measurement on modes 6 and 8. It is evident that, out of these two counts one count is always zero. Alice communicates her PC result to Bob using 2-bit classical channel. However, even after getting Alice’s PC result, Bob cannot construct the original information state. At this instance, the role of the controller, Charlie, becomes crucial. To complete the teleportation protocol, Charlie makes a PC measurement on mode 6 in \{$|EVEN,\alpha\rangle, |ODD,\alpha\rangle$\} basis. The collective PC result in modes ($3,6,8$) can be one of (EVEN/ODD,0,0), (EVEN/ODD,NZE,0), (EVEN/ODD,0,NZE), (EVEN/ODD,ODD,0) and (EVEN/ODD,0,ODD) and we refer to these as cases (1,2), (3,4), (5,6), (7,8) and (9,10) (see Table \ref{table1}). For cases 1 and 2 , the Bob’s state in mode 3 is $|EVEN/ODD,\alpha\rangle$. Since the probabilities for occurrence of any of these cases is, $P_{0}=x^{2}\mbox{cos}^{2}(\theta/2)/(1+x^{2}+x^{4})$ (fig.\ref{Fig:fig-1}) proportional to $|A_{+}|^{2}$, for maximum fidelity, Bob applies unitary transformations $\hat{U}_{0}=|EVEN,\alpha\rangle_{3}\langle{ODD,\alpha}|+|ODD,\alpha\rangle_{3}\langle{EVEN,\alpha}|$ and $\hat{I}$ for cases 1 and 2 respectively, where $\hat{I}$ refers to identity transformation.
\begin{figure}[!htb]
   \begin{minipage}{0.48\textwidth}
     \centering
     \includegraphics[width=.9\linewidth, trim=0 0 0 0, clip]{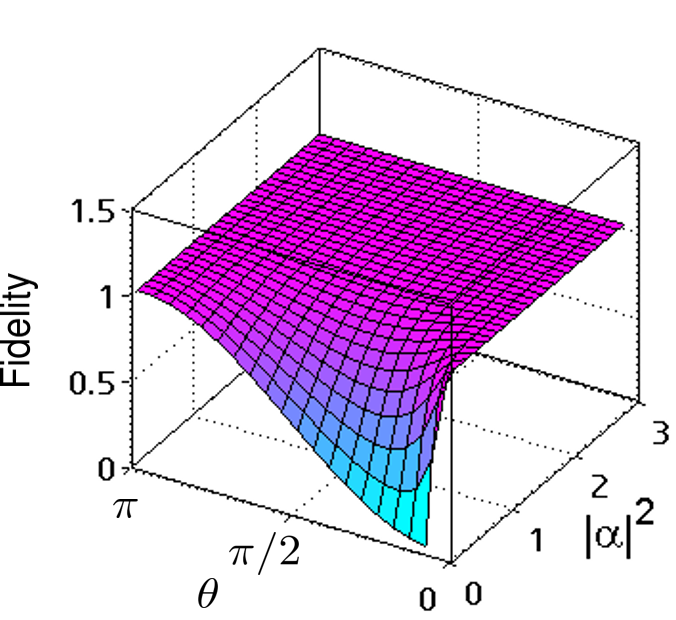}
     \caption{Variation of teleportation fidelity with respect to information parameter $\theta$ and average photon number in the coherent state $|\alpha|^{2}$ for cases 3, 5, 8 and 10.}\label{Fig:fig-3}
   \end{minipage}\hfill
   \begin{minipage}{0.48\textwidth}
     \centering
     \includegraphics[width=.9\linewidth, trim=0 0 0 0, clip]{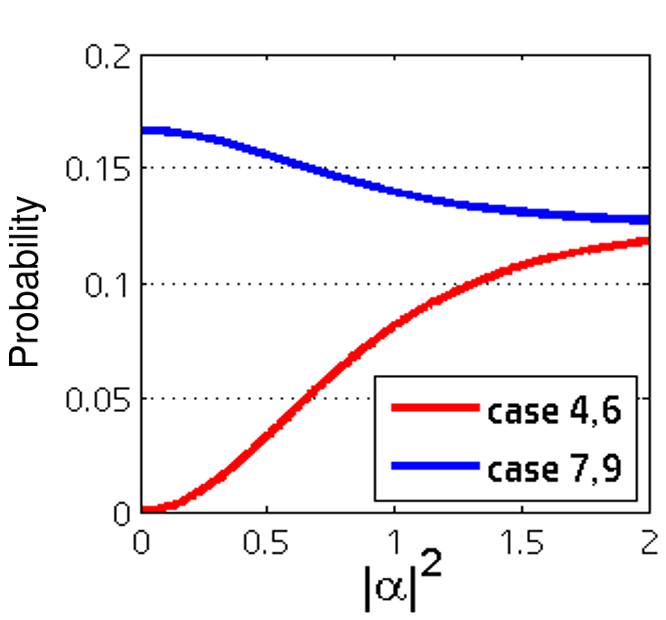}
     \caption{Variation of probability of occurrance with respect to average photon number in the coherent state $|\alpha|^{2}$ for cases 4, 6 (red curve) and for cases 7, 9 (blue curve)}\label{Fig:fig-4}
   \end{minipage}
\end{figure}
For cases 3, 5, 8 and 10, probability for occurrence of each is $P_{1}=(1+x^{4}-2x^{2}\mbox{cos}\theta)/[8(1+x^{2}+x^{4})]$ fig. \ref{Fig:fig-2}. The Bob's states (un-normalized) are $|B_{3}\rangle\sim(1-x^{2})A_{+}|ODD,\alpha\rangle+(1+x^{2})A_{-}|EVEN,\alpha\rangle$, $|B_{5}\rangle\sim(1-x^{2})A_{+}|ODD,\alpha-(1+x^{2})A_{-}|EVEN,\alpha\rangle$, 
$|B_{8}\rangle\sim(1-x^{2})A_{+}|ODD,\alpha+(1+x^{2})A_{-}|EVEN,\alpha\rangle$ and  
$|B_{10}\rangle\sim -(1-x^{2})A_{+}|ODD,\alpha+(1+x^{2})A_{-}|EVEN,\alpha\rangle$
respectively requiring unitary transformations
$\hat{U}_{1}=|EVEN,\alpha\rangle_{3}\langle{ODD,\alpha}|+|ODD,\alpha\rangle_{3}\langle{EVEN,\alpha}|$, 
$\hat{U}_{2}=|EVEN,\alpha\rangle_{3}\langle{ODD,\alpha}|-|ODD,\alpha\rangle_{3}\langle{EVEN,\alpha}|$, 
$\hat{U}_{1}=|EVEN,\alpha\rangle_{3}\langle{ODD,\alpha}|+|ODD,\alpha\rangle_{3}\langle{EVEN,\alpha}|$ and $-\hat{U}_{2}=-|EVEN,\alpha\rangle_{3}\langle{ODD,\alpha}+|ODD,\alpha\rangle_{3}\langle{EVEN,\alpha}|$ respectively, and leading to the same Fidelity, 
$F_{1}=1-[x^{4}\mbox{sin}^{2}\theta/(1+x^{4}-2x^{2}\mbox{cos}\theta)]$ (fig.  \ref{Fig:fig-3}). For cases 4 and 6, probability for occurrence of any case is 
$P_{2}=(1-x^{2})^{2}/[8(1+x^{2}+x^{4})]$
(fig. \ref{Fig:fig-4}, red curve).
Bob’s states are,  
$|B_{4/6}\rangle=A_{+}|EVEN,\alpha\rangle\pm A_{-}|ODD,\alpha\rangle$, requiring unitary transformations $\hat{I}$ and  
$\hat{U}_{3}=|EVEN,\alpha\rangle_{3}\langle{EVEN,\alpha}|-|ODD,\alpha\rangle_{3}\langle{ODD,\alpha}|$, and leading to fidelity $F_{2}=1$. Similarly, for cases 7 and 9, probability for occurrence of each case is 
$P_{3}=(1+x^{2})^{2}/[8(1+x^{2}+x^{4})]$ 
(fig. \ref{Fig:fig-4}, blue curve). Bob's state are $|B_{7/9}\rangle=\pm A_{+}|EVEN,\alpha\rangle+ A_{-}|ODD,\alpha\rangle$, requiring unitary transformations $\hat{I}$ and $-\hat{U}_{3}=-|EVEN,\alpha\rangle_{3}\langle{EVEN,\alpha}|$\\$+|ODD,\alpha\rangle_{3}\langle{ODD,\alpha}|$ respectively, and leading to fidelity $F_{2}=1$.\\
We find that the probabilities converges to a constant value 0.125 for cases 3 to 10 and approaches zero for cases 1 and 2 as mean photons in the coherent states increases. This assures that failure probability is negliglibly small and we can ignore such a case. 
\begin{table}[htb]
\begin{center}
\begin{tabular}{ c c c c c c c c }
\hline 
\label{my-label}
 S. No. & \thead{Alice PC result \\ (modes 5,6)} & \thead{Charlie PC result \\ (mode 3)}  & \thead{Probability} & \thead{Unitary\\ operation} & \thead{Teleported state \\ (unnormalided)} & Fidelity & MAF \\
 \hline
  1 & 0,0  & EVEN & $P_{1}$ & $U_{1}$ & $|T_{0}\rangle$ & $F_{0}$ & 0 \\
 
  2 & 0,0  & ODD & $P_{1}$ & $I$ & $|T_{0}\rangle$ & $F_{0}$ & 0 \\
 
 3 & NZE,0  & EVEN &  $P_{2}$ & $U_{1}$ & $|T_{1}\rangle$ & $F_{1}$ & $1-x^{2}$\\ 
 
 4 & NZE,0  & ODD  & $P_{3}$ & $I$ & $|T_{2}\rangle$ & $1$ & $1$ \\ 
 
 5 & 0,NZE  &  EVEN & $P_{2}$ & $U_{2}$ & $|T_{1}\rangle$ & $F_{1}$ & $1-x^{2}$ \\ 
 
 6 & 0,NZE  & ODD &  $P_{3}$ & $U_{3}$ & $|T_{2}\rangle$ & $1$ & $1$ \\ 
 
 7 & ODD,0  &  EVEN & $P_{4}$ & $I$ & $|T_{2}\rangle$ & $1$ & $1$  \\ 
 
 8 & ODD,0  &  ODD & $P_{2}$ & $U_{1}$ & $|T_{1}\rangle$ & $F_{1}$ & $1-x^{2}$ \\
  
 9 & 0,ODD  &  EVEN & $P_{4}$ & $-U_{3}$ & $|T_{2}\rangle$ & $1$ & $1$ \\
 
 10 & 0,ODD  &  ODD &  $P_{2}$ & $-U_{2}$ & $|T_{1}\rangle$ & $F_{1}$ & $1-x^{2}$ \\ \hline
 
\end{tabular}
\end{center}

\caption{Various possible PC measurement results in modes 6, 8 and 3. For each case the corresponding probability of occurrence, required unitary transformation by Bob, the teleported state, fidelity and MAF is given.}
\label{table1}
\end{table}
\renewcommand{\arraystretch}{2}
\begin{figure}[!htb]
   \begin{minipage}{0.48\textwidth}
     \centering
     \includegraphics[width=.9\linewidth, trim= 0 .5cm 0 0, clip]{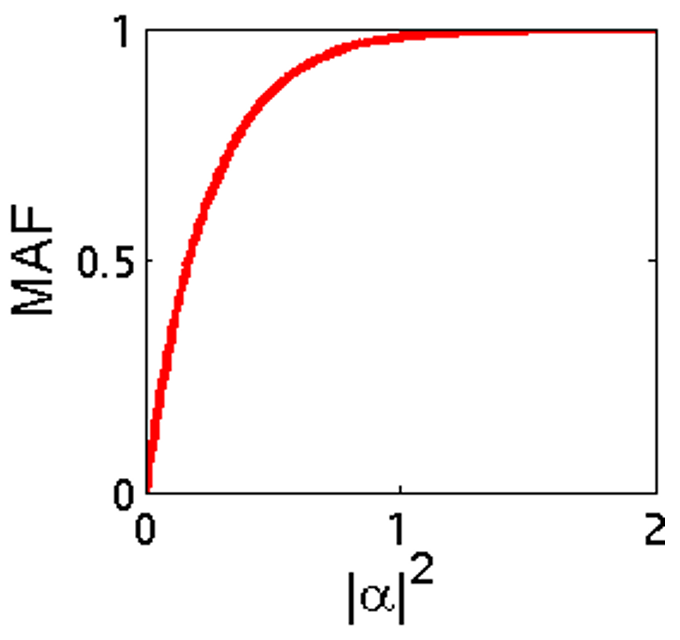}
     \caption{Variation of minimum assured fidelity (MAF) for cases 3, 5, 8 and 10 with respect to average photon number in the coherent state $|\alpha|^{2}$.}\label{Fig:fig-5}
   \end{minipage}\hfill
   \begin{minipage}{0.48\textwidth}
     \centering
     \includegraphics[width=.9\linewidth, trim= 0 0 0 0, clip]{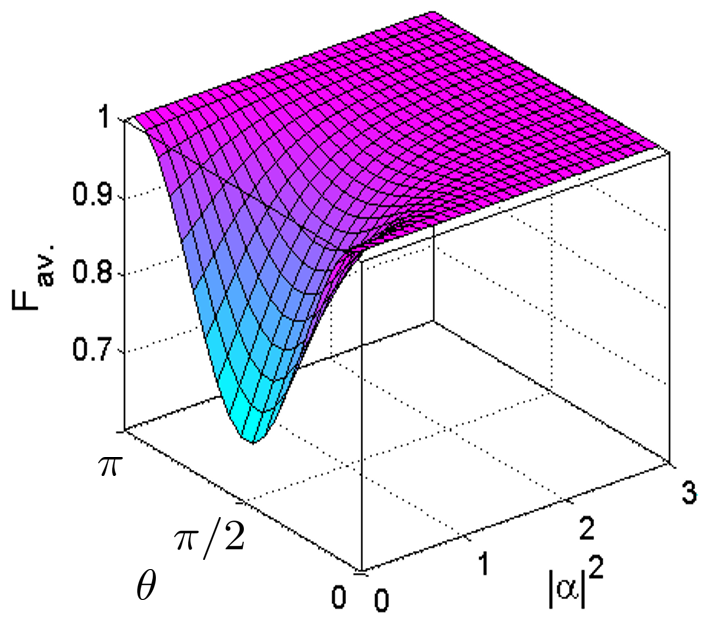}
     \caption{Variation of average  teleportation fidelity with respect to information parameter $\theta$ and average photon number in the coherent state $|\alpha|^{2}$.}\label{Fig:fig-6}
   \end{minipage}
\end{figure} 
\section{Quality of Teleportation}

We note that Fidelities depend on information parameter $\theta$ in many cases. For such a situation Prakash et al. \cite{verma2016standard} defined minimum assured fidelity (MAF) for each case of photon counts and minimum average fidelity ($F_{avg.}$ ) for the overall performance. For cases 1 and 2, the fidelity depends on information and minimum assured fidelity (the minimum for any information) is 0. It has, however, negligible effect on the quality of teleportation as its probability occurrence is almost zero at $|\alpha|^{2}\geq2$.\\
For cases 3, 5, 8 and 10, the fidelity depends on information and minimum assured fidelity (MAF) is $(1-x^{2})$ fig. \ref{Fig:fig-5}.

Despite of the fact that an exact replica of the information state could not be made using any unitary transformation, but exceptionally close state can be achieved. For other cases fidelity is unity and teleportation is ideal. Also, the probabilities converges to a constant value 0.125 for each of the cases 3 to 10 if $|\alpha|^{2}\geq2$ and for all $\theta$ and approaches zero for cases 1 and 2. This assures that failure probability is negligibly small and we can ignore such a case for moderate coherent amplitudes.\\
To estimate overall quality of teleportation, we will calculate average fidelity, defined as the sum of products of probabilities of occurrence with corresponding fidelity for each possible PC result. For our case the average fidelity becomes,

\begin{equation}\label{eq-10}
F_{av.}=\sum_{i=1}^{10}P_{i}F_{i}=1-\dfrac{x^{2}(1+x^{2}\mbox{sin}^{2}\theta)}{2(1+x^{2}+x^{4})}
\end{equation}
with minimum value,
$F_{av. min.}=1-\dfrac{x^{2}(1+x^{2})}{2(1+x^{2}+x^{4})}$
occurring at $\theta=\pi/2$. Fig. \ref{Fig:fig-6} depicts the behaviour of $F_{avg.}$ as a function of $|\alpha|^{2}$ and $\theta$. It is evident $F_{av.}$ converges rapidly and almost become unity for mean photon amplitude as low as 2 and for all $\theta$. Thus, using moderate coherent amplitude an almost perfect CQT of SCS can be achieved.
\section{Conclusion}
Our scheme leads to almost perfect CQT of SCS using 3-mode GHZ type ECS, requiring only linear optical devices such as beam splitter, phase shifters and photon parity detectors. Experimentally, SCS of moderate coherent amplitudes ($\alpha\sim 2$  to  $3$) can be generated in a number of ways such as using Kerr-nonlinear interactions, photon subtraction from squeezed vacuum state, cavity-assisted interactions etc., with high fidelity. Our proposed scheme demands coherent amplitude within this limit which makes our scheme experimentally feasible. In contrast to the previous scheme proposed for CQT of SCS using 4-mode cluster type ECS where the success probability was shown to be 1/2, for our scheme the average fidelity and hence the success almost becomes unity $(\geq.99)$ for moderate mean photons ($|\alpha|^{2}\geq2$) in the coherent state, that too using 3-mode GHZ state. A failure is recorded only when zero counts are obtained in modes 6 and 8. However, as we have already noticed such a case has almost zero probability of occurrence at moderate coherent amplitudes ($|\alpha|^{2}\geq 2 $).\\
Because of the robustness of ECS due to photon absorption, our scheme can be found useful for quantum information processing tasks such as in quantum repeaters and relays, secured long distance quantum communications and also in the recent development of hybrid-quantum information processing, where conversion between discrete and continuous variables is frequently performed. \\

\bibliographystyle{unsrt}
\bibliography{QTrevised}

\end{document}